\documentclass[aps, prd, amsmath, amssymb, floats, floatfix, superscriptaddress, nofootinbib, twocolumn, showpacs,reprint]{revtex4-1}

\usepackage{latexsym}
\usepackage{graphicx}
\usepackage{amsmath}
\usepackage{amssymb}
\usepackage{amsfonts}
\usepackage{times}
\usepackage{xspace} 
\usepackage[usenames]{color}
\usepackage{dcolumn}
\usepackage{bm}
\usepackage{mathrsfs}
\usepackage{tikz}
\usepackage{appendix}
\usepackage{breqn}
\usepackage[]{units}       

\usepackage{subfigure}		
\usepackage[normalem]{ulem} 

\usepackage[tracking=true]{microtype}
\SetTracking{}{500}
\SetTracking{encoding={*}, shape=sc}{40}

\makeatletter
\let\cat@comma@active\@empty
\makeatother

\begin{document} 
 
\title{Stealth black holes in shift symmetric kinetic gravity braiding}

\author{Reginald Christian Bernardo, John Celestial, and Ian Vega}
\affiliation{National Institute of Physics, University of the Philippines, Diliman, Quezon City 1101, Philippines}
\email{rbernardo@nip.upd.edu.ph}
\email{jcelestial@nip.upd.edu.ph}
\email{ivega@nip.upd.edu.ph}
\date{\today}

\begin{abstract}
We derive all hairy stealth black holes in the most general second-order, shift symmetric, scalar-tensor theory with luminally propagating gravitational waves, often called kinetic gravity braiding. Our approach exploits a loophole in a recently obtained no-go statement which claims shift symmetry breaking to be necessary for stealth solutions to exist in kinetic gravity braiding. We highlight the essential role played by a covariantly constant kinetic density in obtaining these solutions. Lastly, we propose a parametrization of the theories based on the asymptotics of its stealth solutions and comment on the intriguing singular effective metric for scalar perturbations in stealth black holes.
\end{abstract}

\maketitle

\section{Introduction}
\label{sec:intro}

There is continuing effort to understand whether the black holes observed by astronomers are indeed the black holes predicted by general relativity. The detection of gravitational waves (GWs) from binary black hole mergers by the LIGO and Virgo collaborations \cite{ligo_catalog} and the recent imaging of the M87 supermassive black hole by the Event Horizon Telescope \cite{eht} draw part of their huge significance from this desire to square new observations with theoretical expectations. They represent our first direct probes of the strong gravity regime, and have provided lampposts (albeit dim) in the otherwise dark landscape of gravitational theories. Further observations by these ongoing efforts, along with future prospective experiments such as the space-based Laser Interferometer Space Antenna (LISA), will greatly inform us of the true nature of black holes and thus help us zero-in on what many hope to be a final theory of gravity \cite{gws_multiband_sesana, gws_multiband_vitale, gws_multiband_giuseppe, gw_astronomy_berti, gw_astronomy_mhz_baibhav, gw_astronomy_multiband_cutler, gw_astronomy_lisa_baker}.

Attendant to this great optimism is the desire to better discriminate between the predictions of general relativity (GR) and alternative theories of gravity. Many alternative theories successfully reproduce the weak-field limits of GR \cite{alternative_gravity_koyama, alternative_gravity_joyce, alternative_gravity_clifton}, and are often far too exceedingly rich in their cosmological phenomenology for them to be usefully constrained by large-scale observations alone \cite{ferreira2019cosmological, constraints_cosmology_li, parametrizing_gravity_clifton, parametrized_lombriser, eft_dark_energy_frusciante}. It is from this context that black hole physics may bring clarity to the prevailing empirical muddle, as theoretical predictions tend to be most different---and thus the most empirically discriminating---in this strong gravity regime \cite{bh_physics_community, scc_cardoso_1, scc_luna_2, superradiance_ghosh, superradiance_ficarra, parametrized_ringdown_cardoso, parametrized_ringdown_mcmanus, strong_gravity_berti, strong_gravity_sathyaprakash}. 

In general relativity, stationary black holes are the epitome of simplicity. They are completely characterized by their mass, electric charge, and spin, according to the celebrated no-hair theorems \cite{no_hair_classic_bekenstein, no_hair_classic_israel, no_hair_classic_israel_2, no_hair_carter}. The new degrees of freedom in alternative theories of gravity threaten to spoil this simplicity, by seemingly supporting black hole solutions that depend on other parameters and that are associated with long-range fields (i.e., ``hair''). However, by some rather surprising twist of irony, various physical considerations often conspire to prevent this from happening \cite{st_black_holes_sotiriou, st_no_hair_theorem_hui, st_no_hair_theorem_sotiriou_2}: the black holes of general relativity also tend to be solutions of alternative theories. In this context, they are called stealth black holes. It also often happens that in some viable alternative theories, these stealth black holes are bald, possessing nothing but trivial (i.e., constant) hair. This has motivated various no-hair extensions to alternative theories of gravity \cite{st_black_holes_sotiriou, st_no_hair_theorem_hui, st_no_hair_theorem_sotiriou_2, st_horndeski_slow_rotation_bh_maselli, antoniou2018evasion}. These theoretical developments are greatly relevant for observations, because they suggest that the strong-field phenomenology of some alternative theories might not be very different from GR, which in turn will make the challenge of observationally discriminating theories even more difficult
\footnote{
For stealth black holes, the perturbative regime is the most promising place for determining smoking-gun signatures of non-GR degrees of freedom \cite{perturbations_eft_tattersall, perturbations_eft_franciolini}.
}.

Our goal in this short paper is to analyze to what extent stealth black hole solutions can be hairy in kinetic gravity braiding (KGB), a phenomenologically interesting class of scalar-tensor theories. Scalar-tensor theories are the oldest and arguably the simplest alternative theories of gravity \cite{alternative_gravity_koyama, alternative_gravity_joyce, alternative_gravity_clifton}. Its inherently wide parameter space makes them desirable for phenomenological purposes, such as for providing mechanisms for inflation and dark energy \cite{st_galileon_inflation_deffayet, st_galileon_inflation_kobayashi, st_galileon_inflation_burrage, eft_de_gleyzes, eft_de_gubitosi, eft_de_cremineli, eft_de_gleyzes2, eft_cosmology_piazza, st_galileon_inflation_kobayashi_2}. The existence of hairy black holes in these theories has also been established in many previous works. For instance, in the most general scalar-tensor theory with second-order field equations, known as Horndeski theories \cite{st_horndeski_seminal, st_horndeski_revive_deffayet_2, st_horndeski_revive_deffayet, st_galileon_inflation_kobayashi_2}, hairy black holes can either be a stealth black hole \cite{st_stealth_minamitsuji_2014, st_stealth_silva, st_horndeski_solutions_babichev_0, st_horndeski_cosmological_tuning_babichev, stealth_ben_achour, stealth_motohashi, st_dhost_bhs_minamitsuji, stealth_minamitsuji, st_kerr_bhs_charmousis}, i.e., GR black hole plus a nontrivial scalar field, or a new hairy black hole \cite{st_horndeski_babichev, st_horndeski_solutions_babichev_2, st_horndeski_cosmological_tuning_babichev, st_horndeski_slow_rotation_bh_maselli, st_horndeski_neutron_stars_maselli, st_horndeski_solutions_kobayashi, st_horndeski_solutions_babichev_0, st_horndeski_solutions_rinaldi, st_horndeski_solutions_anabalon, st_horndeski_solutions_minamitsuji, st_horndeski_solutions_gaete, antoniou2018evasion, antoniou2018black,hairy_bhs_tfl, bernardo2019hair}. There now exists a large and still growing body of literature concerned with the study of novel hairy black holes in Horndeski theories and beyond. See for instance Refs. \cite{st_no_hair_theorem_hui, st_no_hair_benkel, st_no_hair_theorem_sotiriou_1, st_no_hair_theorem_sotiriou_2, st_horndeski_babichev, st_horndeski_solutions_babichev_2, st_horndeski_cosmological_tuning_babichev, st_horndeski_slow_rotation_bh_maselli, st_horndeski_neutron_stars_maselli, st_horndeski_solutions_kobayashi, st_horndeski_solutions_babichev_0, st_horndeski_solutions_rinaldi, st_horndeski_solutions_anabalon, st_horndeski_solutions_minamitsuji, st_horndeski_solutions_gaete, antoniou2018evasion, antoniou2018black,hairy_bhs_tfl, bernardo2019hair} and the references therein.

We shall mostly focus our attention on background solutions in the subset of Horndeski theories called kinetic gravity braiding \cite{st_galileon_inflation_deffayet, st_galileon_inflation_kobayashi}. KGB has become popular lately because of its consistency with the GW170817 bound on the gravitational wave speed \cite{gw_170817_ligo, dark_energy_creminelli, dark_energy_ezquiaga, lombriser2016breaking, lombriser2017challenges} and a number of cosmological constraints \cite{st_horndeski_galileon_barreira_1, st_horndeski_galileon_barreira_2,st_horndeski_renk, st_horndeski_galileon_peirone,peirone2019cosmological,st_horndeski_constraint_mancini2019, horndeski_constraint_noller2018, st_horndeski_constraint_komatsu2019}. Many important theories---quintessence, $k$-essence, cubic Galileon, Galileon ghost condensate, and GR, to name a few---are special cases of KGB (see Eq. (\ref{eq:kgb})). In KGB, various stealth black hole solutions have been obtained in a number of settings and using various methods \cite{stealth_ben_achour, stealth_motohashi, st_dhost_bhs_minamitsuji, hairy_bhs_tfl, stealth_minamitsuji, st_kerr_bhs_charmousis}. For stealth solutions in scalar-tensor theories outside KGB, we refer the reader to Refs. \cite{st_horndeski_solutions_babichev_0, st_stealth_minamitsuji_2014, st_horndeski_cosmological_tuning_babichev, st_stealth_silva, stealth_ben_achour, stealth_motohashi, st_dhost_bhs_minamitsuji, st_kerr_bhs_charmousis}. 

In this paper, we derive stealth black hole solutions with nontrivial hair in shift symmetric KGB.  Previous work \cite{stealth_minamitsuji} has suggested that such solutions are not possible for nonconstant kinetic densities, and that one needs to break shift symmetry in order to accommodate stealth solutions in KGB. However, we uncover a loophole in this no-go result, and then use it to chart an easy path towards obtaining hairy stealth solutions. In fact, we shall derive the most general stealth black hole solutions in shift symmetric KGB.

The rest of this paper proceeds as follows. After a brief review of KGB and its observational constraints (Sec. \ref{sec:kgb}), we proceed by identifying necessary conditions for the existence of stealth hairy black holes in shift symmetric KGB. These conditions locate the sector of shift symmetric KGB where existence is possible. We then derive all hairy stealth black holes in this sector (Sec. \ref{sec:stealth_kgb}), and find that the most general solution essentially consists of a Kerr-(anti) de Sitter metric and a particular class of scalar fields it is able to support. In Sec. \ref{sec:discussion}, we propose a parametrization of the theories based on the asymptotics of its stealth black holes and comment on the intriguing singular effective metric that governs the propagation of scalar modes in stealth solutions. We conclude by addressing future work.

We work with a spacetime with mostly plus metric signature and geometrized units $c = G = 1$.

\section{kinetic gravity braiding theories}
\label{sec:kgb}

Kinetic gravity braiding is described by the action \cite{st_galileon_inflation_deffayet, st_galileon_inflation_kobayashi}
\begin{equation}
\label{eq:kgb}
S_g = \int d^4 x \sqrt{-g} \left[ \kappa R + K \left( \phi, X \right) - G \left( \phi, X \right) \Box \phi \right]
\end{equation}
where $g_{ab}$ is the metric, $R$ is the Ricci scalar, $\kappa = M_P^2/2$ is a bookkeeping constant for the Einstein-Hilbert part of the theory, $\phi$ is the scalar field, $X$ is the scalar field's kinetic density given by $X = - g^{ab} \nabla_a \phi \nabla_b \phi / 2$, and $K$ and $G$ are arbitrary functions of their arguments that we shall refer to as the $k$-essence and braiding potentials, respectively. For brevity, we shall write down $\xi_a =\nabla_a \xi$, $\xi^a = \nabla^a \xi$, and $\xi_{ab} = \nabla_b \nabla_a \xi$ for any scalar function $\xi$, e.g., $X = -g^{ab} \phi_b \phi_b/2$. The term ``braiding" refers to the mixing of scalar ($\psi$) and tensor ($h$) modes present in the term $G \, \Box \phi \sim G \, \partial h \, \partial \psi + O\left( h^2, \psi^2 \right)$ in the second order action which entangles the scalar and tensor modes \cite{st_galileon_inflation_deffayet}. KGB gravity defined by Eq. (\ref{eq:kgb}) is a subset of Horndeski gravity for $G_2 = K$, $G_3 = G$, $G_4 = \kappa$, and $G_5 = 0$ for arbitrary functions $K$ and $G$ \cite{st_dark_energy_tsujikawa, horndeski_review_kobayashi2019}. On the other hand, $k$-essence is a special case of KGB for constant $G$ \cite{kessence_seminal_armendariz, kessence_seminal_armendariz2}. Scalar-tensor theories can also be invariant with respect to the shift transformation, $\phi \rightarrow \phi \, + \, k$ for constant $k$. In the context of KGB, shift symmetry means that $K = K \left( X \right)$ and $G = G \left( X \right)$. The cubic Galileon is a special case of shift symmetric KGB for $K \sim X$ and $G \sim X$ as well as the Galileon ghost condensate for $K \sim X + X^2$ and $G \sim X$ \cite{st_dark_energy_tsujikawa}.

KGB is among the few alternative gravity theories favored by the stringent GW170817 constraint restricting GWs to propagate very nearly at the speed of light \cite{gw_170817_ligo, dark_energy_creminelli, dark_energy_ezquiaga, lombriser2016breaking, lombriser2017challenges}. Unlike many of its competitors, KGB has only tensor modes that propagate on the light cone. Shift symmetric KGB is, thus, the most general, second-order, shift symmetric scalar-tensor theory with luminally propagating gravitational waves. From the effective field theory (EFT) standpoint, GWs can also depend on the frequency, and so
the event GW170817 can be regarded as valid only for the LIGO frequency band ($\sim 100$ Hz) \cite{dark_energy_derham}. A GW detection with an optical counterpart in LISA (0.1 mHz - 0.1 Hz) will put a stronger conclusion to the GW speed. On the other hand, several works have shown that the cubic Galileon is disfavored by cosmic microwave background, baryon acoustic oscillations, integrated Sachs-Wolfe, and weak lensing \cite{st_horndeski_galileon_barreira_1, st_horndeski_galileon_barreira_2,st_horndeski_renk, st_horndeski_galileon_peirone} and, recently, it has been claimed that cosmological data favor the Galileon ghost condensate over $\Lambda$CDM \cite{peirone2019cosmological}. Also, several recent works have shown that cosmological data cannot rule out KGB \cite{st_horndeski_constraint_mancini2019, horndeski_constraint_noller2018, st_horndeski_constraint_komatsu2019} and that the braiding might even be the inevitable limit of scalar-tensor theories should all future cosmological data continue to remain consistent with GR \cite{no_slip_linder, no_slip_cmb_brush, no_run_linder}. Lastly, viable alternative theories must possess screening mechanisms to pass solar system tests \cite{screening_chakraborty, screening_brax, screening_schmidt, screening_davis, st_horndeski_vainshtein_dima, screening_mcmanus}. The existence of the Vainshtein mechanism is well-known in the cubic Galileon theory but whether screening mechanisms exist for the rest of KGB is still an open question \footnote{
In general, the analysis of screening mechanisms boils down to first specifying the theory's action and then determining whether nonlinear interactions dominate when the coupling constants take on their local limits \cite{screening_mcmanus}.
}.

The variation of the shift symmetric KGB action with respect to the metric and the scalar field, respectively, leads to the vacuum field equations
\begin{equation}
\label{eq:bg_mfe} 
\begin{split}
& \kappa G_{ab} - \frac{1}{2} g_{ab} K - \frac{1}{2} \phi_a \phi_b K_{X} \\
&+ \left[ \frac{1}{2} \phi_a \phi_b \Box \phi - \phi_{( a} \phi_{b) c} \phi^c + \frac{1}{2} g_{ab} \phi^c \phi^d \phi_{cd} \right] G_X = 0
\end{split}
\end{equation}
and
\begin{equation}
\begin{split}
\label{eq:bg_sfe} 
& \Box \phi K_{X} - \phi^a \phi^b \phi_{ab} K_{XX} \\
& + \left[ - \phi^a \nabla_a \Box \phi - \left( \Box \phi \right)^2 + \phi^a \Box \phi_a + \phi_{ab} \phi^{ab} \right] G_X \\
& \phantom{avengers assemble} + \phi^b \phi_{ab} \left[ \phi^a \Box \phi - \phi^d \phi_d^{\ a} \right] G_{XX} = 0 
\end{split}
\end{equation}
where a subscript of $X$ in the potentials $K$ and $G$ means differentiation with respect to $X$, e.g., $K_{XX} = d^2K/dX^2$, and the symmetrization rule for a tensor $T_{ab}$ is $T_{(ab)} = \left( T_{ab} + T_{ba} \right) /2$. Eq. (\ref{eq:bg_mfe}) is the Einstein equation with a stress-energy tensor (SET) $T^{(\phi)}_{ab}$ given by
\begin{equation}
\begin{split}
8 \pi \kappa T^{(\phi)}_{ab} = & \ g_{ab} \frac{K}{2} + \phi_a \phi_b \frac{K_{X}}{2} \\
& - \bigg[ \phi_a \phi_b \Box \phi - 2 \phi_{( a} \phi_{b) c} \phi^c + g_{ab} \phi^c \phi^d \phi_{cd} \bigg] \frac{ G_X }{2} .
\end{split}
\end{equation}
On the other hand, Eq. (\ref{eq:bg_sfe}) is the equation of motion of the scalar field. A solution $\left(g_{ab}, \phi \right)$ of kinetic gravity braiding $(K, G)$ must simultaneously satisfy Eqs. (\ref{eq:bg_mfe}) and (\ref{eq:bg_sfe}). 

\section{Stealth black holes in kinetic gravity braiding}
\label{sec:stealth_kgb}

In this section, we expose a loophole in the no-go theorem obstructing the existence of stealth solutions in shift symmetric KGB (Sec. \ref{subsec:no_go_loophole}). We take advantage of this loophole to determine conditions that will instead identify the specific sector of shift symmetric KGB in which stealth solutions exist (Sec. \ref{subsec:hairy_stealth_kgb}).  For concreteness, we explicitly construct the stealth Kerr-(anti) de Sitter black hole solution with linearly time dependent scalar field (Sec. \ref{subsec:kerr_ds}). The most general solution is a Kerr-(anti) de Sitter metric plus a scalar field with constant kinetic density.

\subsection{A stealthy loophole}
\label{subsec:no_go_loophole}

The ground work for building stealth solutions was laid out in Ref.~\cite{stealth_minamitsuji}. This important paper also derived stealth Schwarzschild black hole solutions in KGB for the first time. In doing so, the authors of Ref. \cite{stealth_minamitsuji} argued that for nontrivial stealth solutions $\left( \phi_a \neq 0 , \nabla_a X \neq 0 \right)$ to exist, either shift symmetry must be broken, $K = K\left( \phi, X \right)$ and $G = G\left( \phi, X \right)$, or else the KGB must be trivial. Their proof starts with necessary conditions for obtaining stealth solutions \cite{stealth_minamitsuji}: 
\begin{eqnarray}
\label{eq:stealth_minamitsuji_k} K_X \left( \phi_0, X_0 \right) &=& 0 \\
\label{eq:stealth_minamitsuji_g} G_X \left( \phi_0, X_0 \right) &=& 0 
\end{eqnarray}
where $\phi_0$ and $X_0$ are the background scalar field and kinetic density, respectively. For shift symmetric KGB, $K = K\left(X\right)$ and $G = G\left(X\right)$, the above necessary conditions successively lead to $K_X\left(X_0\right) = K_{XX}\left(X_0\right) = K_{XXX}\left(X_0 \right) = \cdots = 0$ and $G_X\left(X_0\right) = G_{XX}\left(X_0\right) = G_{XXX}\left(X_0 \right) = \cdots = 0$ \emph{provided} that $\nabla_a X_0 \neq 0$. The no-go theorem then states that KGB with stealth solutions must go beyond shift symmetry; otherwise, the KGB must be trivial.

The foregoing no-go statement is correct. But it is anchored on the important provision that $\nabla_a X_0 \neq 0$. When the kinetic density is covariantly constant, i.e., $\nabla_a X_0 = 0$, the no-go statement no longer holds, and the tension between shift symmetry and stealth solutions disappears. In fact, we shall assert an even stronger statement: for stealth solutions to exist in shift symmetric KGB, the kinetic density needs to be covariantly constant. We prove this in the next section\footnote{This result should be viewed as a complement to the no-go statement of Ref. \cite{stealth_minamitsuji}, which focuses on stealth solutions with $\nabla_a X_0 \neq 0$.}.

\subsection{Conditions for hairy stealth black holes}
\label{subsec:hairy_stealth_kgb}

To obtain stealth black hole solutions, we require that the field equations (Eqs. \eqref{eq:bg_mfe} and \eqref{eq:bg_sfe}) reduce to the Einstein equation with a cosmological constant $\Lambda$:
\begin{equation}
\label{eq:einstein_cc}
G_{ab} + \Lambda g_{ab} = 0 .
\end{equation}
Reducing Eqs. \eqref{eq:bg_mfe} and \eqref{eq:bg_sfe} to Eq. \eqref{eq:einstein_cc} assures that the KGB field equations are satisfied by the same black hole geometries of GR, i.e., Kerr-(anti) de Sitter black hole, although with a nontrivial scalar field.

The equivalence of Eq. (\ref{eq:einstein_cc}) with Eq. (\ref{eq:bg_mfe}) demands
\begin{equation}
\label{eq:condition_stealth_kgb}
\begin{split}
\Lambda g_{ab} = 
& - g_{ab} \frac{ K}{2\kappa} - \phi_a \phi_b \frac{ K_{X} }{2\kappa} \\
& + \left[ \phi_a \phi_b \Box \phi - 2 \phi_{( a} \phi_{b) c} \phi^c +  g_{ab} \phi^c \phi^d \phi_{cd} \right] \frac{G_X}{2 \kappa} .
\end{split}
\end{equation}
Then, taking the trace and the contraction with $\phi^a \phi^b$ of Eq. \eqref{eq:condition_stealth_kgb} yields
	\begin{equation}
		\label{eq:condition_stealth_kgb_trace}
		2K - X K_X + 4 \kappa \Lambda + \left( X \Box \phi - \phi^c \phi^d \phi_{cd} \right) G_X = 0
	\end{equation}
and
	\begin{equation}
		\label{eq:condition_stealth_kgb_pp}
		K - 2 X K_X + 2 \kappa \Lambda + \left( 2 X \Box \phi + \phi^c \phi^d \phi_{cd} \right) G_X = 0,
	\end{equation}
respectively. These are the necessary conditions for obtaining stealth solutions in KGB. 

Isolating $K_X$ and $G_X$ from Eqs. \eqref{eq:condition_stealth_kgb_trace} and \eqref{eq:condition_stealth_kgb_pp} to get the functionals
\begin{eqnarray}
\label{eq:K_X_constraint} 
K_X [K,\Lambda] &=& \left( K + 2 \kappa \Lambda \right) \left( \dfrac{\Box \phi}{\phi^c \phi^d \phi_{cd}} + \dfrac{1}{X} \right) \\
\label{eq:G_X_constraint} 
G_X [K,\Lambda] &=& \dfrac{ K + 2 \kappa \Lambda }{ \phi^c \phi^d \phi_{cd} },
\end{eqnarray}
and substituting these expressions back into Eq. \eqref{eq:condition_stealth_kgb} gives 
\begin{equation}
\label{eq:branches}
- \left( \dfrac{K}{2\kappa} + \Lambda \right) \left( \dfrac{ \phi_c \phi_d \phi^a \phi^b \phi_{ab} + 2X \phi^a \phi_{a(c}\phi_{d)} }{ X \left( \phi^i \phi^j \phi_{ij} \right) } \right) = 0 .
\end{equation}
Clearly, it is possible to factor out the vector $\phi^a \phi_{ab}$ from the numerator, and so Eq. \eqref{eq:branches} can be written as
\begin{equation}
\label{eq:branches_2}
- \left( \dfrac{K \left( X_0 \right) }{2\kappa} + \Lambda \right) \left( \phi^b \phi_c \phi_d + 2 X \delta^b_{\ ( c} \phi_{d)} \right) \left( \dfrac{ \nabla_b \ln X }{ \phi^j \nabla_j X} \right) = 0
\end{equation}
noting that
\begin{equation}
\label{eq:grad_X}
\nabla_b X = - \phi^a \phi_{ab} .
\end{equation}
Here we are now displaying an explicit dependence on $X_0$ to emphasize that it is evaluated on the background kinetic density.
Eq.~\eqref{eq:branches_2} is a condition that must be satisfied by the scalar field in order for a GR black hole geometry, described by Eq. \eqref{eq:einstein_cc}, to be dressed with hair. In what follows, we unpack this condition to understand its implications. 

First off, Eq.~\eqref{eq:branches_2} implies a covariantly constant kinetic density. To see this, we first note that the equation is satisfied in three possible ways, corresponding to three branches of solutions: the first branch
\begin{equation}
\label{eq:branch1_kgb}
K \left( X_0 \right) = - 2\kappa \Lambda ,
\end{equation}
the second branch
\begin{equation}
\label{eq:branch2_kgb}
\phi^b \phi_c \phi_d + 2 X \delta^b_{\ ( c} \phi_{d)} = 0 ,
\end{equation}
and the third branch
\begin{equation}
\label{eq:branch3_kgb}
\dfrac{ \nabla_b \ln X }{ \phi^j \nabla_j X} = 0 .
\end{equation}

The first branch (Eq.~\eqref{eq:branch1_kgb}) requires the background kinetic density $X_0$ to be constant. If $X_0 = X_0(x)$ possessed some nontrivial dependence on the spacetime coordinates $x^a$, then because $K[X]$ is a nontrivial functional of its argument \footnote{KGB requires that its potentials $K=K[X]$ and $G=G[X]$ be nontrivial functions of $X$, otherwise KGB just reduces to GR.}, $K[X_0(x)] = K(x)$ will also be a non-constant function in spacetime. Eq.~\eqref{eq:branch1_kgb} and the nontriviality of $K[X]$ force $X_0$ to be covariantly constant.

As a consequence, the KGB potentials and their derivatives also become constants on the background. Substituting Eq. \eqref{eq:branch1_kgb} back into Eqs. \eqref{eq:condition_stealth_kgb_trace} and \eqref{eq:condition_stealth_kgb_pp} leads to
\begin{equation}
\label{eq:branch1_final_recast}
K_X \Box \phi = G_X .
\end{equation}
For shift symmetric solutions, this can be satisfied only when $K_X = 0, G_X = 0$, and $\Box \phi \neq 0$ \footnote{We disregard the case $\Box \phi = 0$, for which the scalar is massless and minimally coupled to the black hole, as it is well known that such scalar-tensor theories are subject to the no-hair theorem of \cite{st_black_holes_sotiriou}.}. To see this, suppose that $K_X \neq 0$ and $G_X \neq 0$. Then, Eq. \eqref{eq:branch1_final_recast} can be read as the dynamical equation for a scalar field $\phi$ that is not shift symmetric. More concretely, Eq. \eqref{eq:branch1_final_recast} is the field equation of the shift symmetry breaking theory, $L = X + \left( G_X / K_X \right) \phi $, for a massive minimally coupled scalar field $\phi$ with a constant scalar charge $G_X/K_X$. In the first branch, the background $K_X$ and $G_X$ must then necessarily vanish. The demand for stealth solutions in shift-symmetric KGB therefore limits it to the $(K, G)$-theory space defined by 
\begin{equation}
\label{eq:option2_kgb}
K \left( X_0 \right) = -2\kappa \Lambda \ \ \ \ \text{and} \ \ \ \ K_X \left( X_0 \right) = G_X \left( X_0 \right) = 0 
\end{equation}
for some constant $X_0$.

Now consider the second branch. Contracting the indices $b$ and $c$ of Eq. \eqref{eq:branch2_kgb}, we then obtain
\begin{equation}
\label{eq:branch2_necessary}
X \phi_d = 0 
\end{equation}
as a necessary condition. This shows that, in the second branch, the solutions are just the trivial solution $\phi_a = 0$ or a vanishing kinetic density $X = 0$, both of which pertain to a vanishing background scalar field stress-energy tensor. As we are only after hairy stealth solutions, the second branch is of no interest.

Finally, we move to the third branch. Contracting it with $\phi^b$, we obtain
\begin{equation}
\label{eq:branch3_necessary}
\dfrac{1}{X} = 0 .
\end{equation}
This shows that the background kinetic density blows up in the third branch, i.e., the stress-energy tensor of the scalar field diverges, and is of no physical interest.

Clearly then, the first branch (Eq. \eqref{eq:branch1_kgb}) is the only viable option, and again, this amounts to the requirement that the background kinetic density is covariantly constant. 

Now, to be able to claim that we have obtained a solution to the KGB field equations, we must also show that the scalar field equation (Eq. \eqref{eq:bg_sfe}) is identically satisfied. Using Eq. \eqref{eq:grad_X}, we find that a constant kinetic density also means
\begin{equation}
\label{eq:constant_X_also}
\phi^a \phi_{ab} = 0 .
\end{equation}
The second and fourth terms of Eq.~\eqref{eq:bg_sfe} vanish as a result. The first and third terms, on the other hand, already vanish due to the constraints set by Eq. \eqref{eq:option2_kgb}. Yet again, we see the importance of a constant kinetic density, this time for satisfying the scalar field equation.

To summarize, we see that after imposing Eqs. (\ref{eq:option2_kgb}) and (\ref{eq:constant_X_also}), the field equations of KGB (Eqs. \eqref{eq:bg_mfe} and \eqref{eq:bg_sfe}) finally reduce to Eq. (\ref{eq:einstein_cc}). With this, KGB is able to accommodate GR's black hole geometries with nontrivial scalar hair, $\phi_a \neq 0$. More important for what follows, we find that stealth black holes in shift symmetric KGB are restricted to the $(K, G)$-sector defined by
\begin{eqnarray}
\label{eq:constraint_k}
K \left( X_0 \right) &=& - 2 \kappa \Lambda \\
\label{eq:constraint_kxgx}
K_X \left( X_0 \right) &=& G_X \left( X_0 \right) = 0
\end{eqnarray}
for some constant $X_0$, defining the scalar hair's background kinetic density \footnote{This set of constraints can be viewed as a special limit of analogous investigation in DHOST (See Ref. \cite{stealth_motohashi}). We emphasize, however, that in this paper the result was derived solely in shift symmetric KGB and therefore complements the work of Ref. \cite{stealth_minamitsuji}.}.

\subsection{Kerr-(anti) de Sitter black hole with linearly time dependent scalar hair}
\label{subsec:kerr_ds}

We now make use of the results of the previous section by showing explicitly the \textit{special case} of a rotating black hole solution (Kerr-(anti) de Sitter) with linearly time dependent hair in shift symmetric KGB. The interest in this cannot be overstated as rotating black holes are ubiquitous in nature and cosmological data suggest that we live in a universe with a non-vanishing cosmological constant $\Lambda$.

The most general rotating black hole solution to Eq. \eqref{eq:einstein_cc} is given by \cite{st_kerr_bhs_charmousis}
\begin{equation}
\label{eq:kerr_ds}
\begin{split}
ds^2 = 
& - \dfrac{ \Delta_r }{ \Xi^2 \rho^2 } \left( dt - a \sin^2 \theta d \varphi \right)^2 + \rho^2 \left( \dfrac{ dr^2 }{ \Delta_r } + \dfrac{ d\theta^2 }{ \Delta_\theta } \right) \\
& + \dfrac{ \Delta_\theta \sin^2 \theta }{ \Xi^2 \rho^2 } \left[ a dt - \left( r^2 + a^2 \right) d \varphi \right]^2
\end{split}
\end{equation}
where
\begin{eqnarray}
\Delta_r &=& \left( 1 - \dfrac{r^2}{l^2} \right) \left( r^2 + a^2 \right) - 2Mr \\
\Delta_\theta &=& 1 + \dfrac{a^2}{l^2} \cos^2 \theta \\
\Xi &=& 1 + \dfrac{a^2}{l^2} \\
\rho^2 &=& r^2 + a^2 \cos^2 \theta \\
l &=& \sqrt{ \dfrac{3}{\Lambda} } .
\end{eqnarray}
It is easy to check that Eq.~\eqref{eq:kerr_ds} reduces to the Kerr solution as $\Lambda \rightarrow 0$ ($l \rightarrow \infty$). We now dress this with a linearly time dependent scalar field
\begin{equation}
\label{eq:scalar_linear_time}
\phi = q t + \phi_0 \left( r , \theta \right)
\end{equation}
where $q$ is a constant and $\phi_0$ is a function only of the radial coordinate $r$ and polar coordinate $\theta$. The motivation for time dependence is to break the staticity assumption of the no-hair theorem for the Galileon \cite{st_horndeski_solutions_babichev_0, st_no_hair_theorem_hui, st_horndeski_babichev}. We can easily recover the stationary solution by setting $q=0$. 

To see the viability of a stealth Kerr-(anti) de Sitter solution, we explicitly derive the gradient of the scalar field. We consider the most general axisymmetric scalar field with linear time dependence
\begin{equation}
\label{eq:scalar_axisymmetric_q}
\phi = q t + \phi_0 \left( r , \theta \right).
\end{equation}
Assuming a sum-separable ansatz
\begin{equation}
\label{eq:sum_separable}
\phi_0 = \chi \left( r \right) + \psi \left( \theta \right)
\end{equation}
for the axisymmetric piece, we then obtain the $\phi_b$-components
\begin{align}
\label{eq:scalar1_kerr_ds} 
\left( \frac{d \chi}{dr} \right)^2 
&= \dfrac{2 \alpha^2 r^2 - C }{\Delta_r} + q^2 \Xi^2 \frac{ \left(r^2+a^2\right)^2 }{\Delta_r^2} \\
\label{eq:scalar2_kerr_ds} 
\left( \frac{d \psi}{d\theta} \right)^2 
&= \dfrac{ 2 \alpha^2 a^2 \cos^2 \theta + C }{\Delta_\theta} - \left( \dfrac{ q a \Xi \sin \theta }{ \Delta_\theta } \right)^2
\end{align}
by solving $X = -\alpha^2$ for some constant $\alpha$ \footnote{The minus sign in $X_0 = - \alpha^2$ is motivated in the Schwarzschild limit, where the background kinetic density becomes $X_0 = q^2/2f - f \phi_0^{\prime 2} / 2$. Outside the event horizon and for large $\phi_0'$, the kinetic density is negative.}. In Eqs. \eqref{eq:scalar1_kerr_ds} and \eqref{eq:scalar2_kerr_ds}, $C$ is a separation constant. As we are working in shift symmetric theory, Eqs. \eqref{eq:scalar1_kerr_ds} and \eqref{eq:scalar2_kerr_ds} and the Kerr-(anti) de Sitter metric (Eq. \eqref{eq:kerr_ds}) constitute a solution to KGB in the $(K, G)$-space defined by $K \left( - \alpha^2 \right) = -2 \Lambda \kappa$ and $K_X \left( - \alpha^2 \right) = G_X \left( -\alpha^2 \right) = 0$ for some constant $\alpha$ which sets the background kinetic density.

\section{Discussion}
\label{sec:discussion}

The analysis provided in Sec.~\ref{subsec:hairy_stealth_kgb} differs significantly from that of Ref. \cite{stealth_minamitsuji}, and it serves to better highlight the importance of the constancy of the kinetic density as a condition for stealth solutions. Specifically, we have shown that in shift symmetric KGB, the kinetic density must necessarily be constant for stealth solutions to even exist. We view this as an important complement to the results of Ref. \cite{stealth_minamitsuji}, which applies to a broader class of theories. 

Our analysis selects the particular sector of KGB theories that admits hairy stealth solutions. It is prudent to try to locate this sector within the broader scheme of shift-symmetric Horndeski theories. One classification scheme for shift symmetric Horndeski theories was introduced in Ref. \cite{st_no_hair_theorem_sotiriou_1} based on the no-hair theorem for the Galileon \cite{st_no_hair_theorem_hui}. In this classification scheme, Horndeski theories can either accept the trivial solution ($\phi' = 0$ and $g_{ab}=$ Schwarzschild) or accommodate hairy black holes (via a coupling with the Gauss-Bonnet scalar or through singular Horndeski potentials). The stealth theories of Sec. \ref{sec:stealth_kgb} belong to the former class if the potentials are nonsingular but to the latter class if the potentials are singular. If the potentials are nonsingular, then the trivial solution ($\phi' = 0$ and $g_{ab}=$ Schwarzschild) can be obtained. For singular potentials, the $\phi' = 0$ limit is excluded as a formal solution and so the black holes must necessarily have hair. A more relaxed classification scheme recently introduced in Ref. \cite{st_horndeski_classification_saravani} is based on shift symmetric Horndeski theories' consistency with the GR limit or, at the very least, Lorentz invariance for $\phi' = 0$. The fairly general constraints on the space of KGB (Eq. \eqref{eq:option2_kgb}) with stealth solutions show that the KGB of Sec. \ref{sec:stealth_kgb} can either satisfy Lorentz invariance or violate it in the limit $\phi' = 0$. Lorentz invariance will be violated if the KGB potentials and/or their derivatives blow up in the limit $\phi' = 0$. 

For the asymptotically (anti) de Sitter stealth black holes, we can parametrize the KGB potentials as
\begin{equation}
\label{eq:potentials_stealth} 
K(X) = F(X) + 2 \alpha \frac{ F' \left( -\alpha^2 \right) }{ H' \left( \alpha \right) } H \left( \sqrt{-X} \right) 
\end{equation}
and
\begin{equation}
\label{eq:potentials_cubic_stealth} 
G(X) = P(X) + 2 \alpha \frac{ P' \left( -\alpha^2 \right) }{ Q' \left( \alpha \right) } Q \left( \sqrt{-X} \right) 
\end{equation}
where $F$, $H$, $P$, and $Q$ are continuously differentiable but otherwise arbitrary functions. A special case of this parametrization is the stealth SAdS black hole of Ref. \cite{hairy_bhs_tfl} which corresponds to $F(x) = x$, $H(x) = x$, constant $P(x)$ and $Q(x)$, and $\alpha = \beta$. The parametrization given by Eqs. (\ref{eq:potentials_stealth}) and (\ref{eq:potentials_cubic_stealth}) conveniently distinguishes solutions based on their asymptotics at infinity. For asymptotically AdS black holes,
\begin{equation}
F \left( - \alpha^2 \right) + 2 \alpha \frac{ F' \left( -\alpha^2 \right) }{ H' \left( \alpha \right) } H \left( \alpha \right) > 0  ,
\end{equation}
and for asymptotically dS black holes,
\begin{equation}
F \left( - \alpha^2 \right) + 2 \alpha \frac{ F' \left( -\alpha^2 \right) }{ H' \left( \alpha \right) } H \left( \alpha \right) < 0  .
\end{equation}
The arbitrariness of $F$, $H$, $P$, and $Q$ guarantees that both cases can indeed be satisfied, e.g., $F(x) = -x$, $H(x) = x$, and arbitrary $P(x)$ and $Q(x)$ lead to asymptotically dS stealth black holes. Some choices also lead to asymptotically flat solutions, e.g., $F(x) = x$, $H(x) = -x^2$, arbitrary $P(x)$ and $Q(x)$. In such cases, both the $k$-essence potential $K$ and its derivative $K_X$ vanish on the background. For KGB with asymptotically flat stealth, Schwarzschild or Kerr, black holes, we can instead parametrize the $k$-essence potential as
\begin{equation}
\label{eq:potentials_stealth_schw}
\begin{split}
K(X) = & \ \gamma \ln \left( \frac{F \left( X \right)}{F\left(-\alpha^2 \right)} \right) \\
& + 2 \alpha \gamma \frac{ H\left( \alpha \right) F'\left( -\alpha^2 \right) }{ H'\left(\alpha\right) F\left(-\alpha^2\right) } \ln \left( \frac{ H \left( \sqrt{-X} \right) }{ H\left(\alpha\right) } \right) 
\end{split}
\end{equation}
where $\gamma$ is a constant. It is easy to see that Eq. (\ref{eq:potentials_stealth_schw}) guarantees that $K \left( - \alpha^2 \right) = 0$ and $K_X \left( - \alpha^2 \right) = 0$ on the background $\left( g_{ab} = \text{Kerr}, X = - \alpha^2 \right)$. We point out that the asymptotic behavior of the spacetime at infinity is determined only by the background value of the $k$-essence potential. 

We have yet to comment on the behavior of perturbations in this background. This question is deeply connected with the stability of solutions and the physical nature of the new degrees of freedom in alternative theories. For scalar-tensor theories, some insight can be gained from the effective metric or the causal cones governing the propagation of gravitational perturbations \cite{stability_babichev, stability_esposito}. For example, in $k$-essence, the causal cone for scalar modes is described by the effective metric
\begin{equation}
S^{ab} = K_X g^{ab} - K_{XX} \phi^a \phi^b
\end{equation}
while the causal cone for the tensor degrees of freedom is the light cone of $g^{ab}$. For our stealth solutions, the effective metric for scalar modes becomes
\begin{equation}
S^{ab}_{\text{stealth}} = - K_{XX} \phi^a \phi^b .
\end{equation}
As first pointed out in Ref. \cite{st_bh_pt_de_rham}, the scalar-mode effective metric is singular on stealth black holes \footnote{Ref. \cite{st_bh_pt_de_rham}, as well as Ref. \cite{st_kerr_pt_charmousis}, came out at the time of writing of this manuscript.}. Already, this puts into question the viability of the theory as an EFT \footnote{A potential solution to this which relies on degeneracy conditions has recently been given in Ref. \cite{scordatura_motohashi}.}. Outside this EFT perspective, the singular metric implies that the scalar modes are nondynamical, and that therefore the only propagating degrees of freedom are the transverse-traceless pieces of the tensor perturbations $h_{ab}$. A quick calculation of the infinite sound speed for scalar modes supports this interpretation
\cite{perturbation_kobayashi_1, perturbation_kobayashi_2, stealth_minamitsuji}. This nondynamicality of the scalar modes also extends to the stealth black holes of KGB. By expanding the KGB action up to second order in the perturbations, it can be shown that the effective metric for the scalar perturbations in a stealth black hole is \cite{st_galileon_inflation_deffayet}
\begin{equation}
\label{eq:effective_metric_kgb}
S^{ab}_{\text{stealth}} \sim \phi^a \phi^b.
\end{equation}
The effective metric given by Eq. (\ref{eq:effective_metric_kgb}) leads to a nonhyperbolic equation (nonpropagating) for the scalar modes. As an example, for a Schwarzschild-(anti) de Sitter black hole (described by the metric function $f = 1 - 2M/r - \Lambda r^2/3$) with linearly time dependent scalar field we have
\begin{equation}
\label{eq:effective_metric_kgb_example}
\phi^a \phi^b
= \left(
\begin{array}{cccc}
\dfrac{q^2}{f^2} & -q \phi_0' & \phantom{gg} 0 \phantom{g} & \phantom{g} 0 \phantom{gg} \\
-q \phi_0' & f^2 \phi_0^{\prime 2} & \phantom{gg} 0 \phantom{g} & \phantom{g} 0 \phantom{gg}  \\ 
0 & 0 & \phantom{gg} 0 \phantom{g} & \phantom{g} 0 \phantom{gg}\\
0 & 0 & \phantom{gg} 0 \phantom{g} & \phantom{g} 0 \phantom{gg}\end{array}
\right),
\end{equation}
which is obviously singular (its determinant vanishes). The implications of this on the scalar modes will be further explored in a forthcoming paper \cite{perturbations_bernardo}. One might think that this conclusion stems from the constancy condition imposed on the kinetic density \cite{st_bh_pt_de_rham}. However, in Ref. \cite{stealth_minamitsuji}, we see that this same conclusion is reached even for \emph{non-constant} kinetic densities and for shift symmetry breaking KGB, where $K_X \left(\phi_0, X_0\right) = 0$ and $G_X \left(\phi_0, X_0\right) = 0$ also emerge as necessary conditions for stealth solutions (with $\phi_0$ as the background scalar field and $X_0$ as a non-constant kinetic density). It is thus tempting to speculate that the effective metric for the scalar modes on stealth black holes may be singular in general.

In light of these considerations, it is of great physical interest to understand how modified gravity manifests in stealth black holes, particularly when the additional degrees of freedom are nondynamical. We suspect that the master equation for tensor perturbations is modified only by an effective source term, and that this may have important observational implications for future space-based gravitational wave missions such as LISA. Very recently, this was shown to be the case for the perturbations of a stealth Kerr black hole in DHOST theories \cite{st_kerr_pt_charmousis}. Our independent analysis of gravitational perturbations in nonrotating stealth black holes in KGB also supports this and will be discussed in a forthcoming paper \cite{perturbations_bernardo}.

\section{Conclusions}
\label{sec:conclusions}

This paper spells out general conditions for the existence of hairy stealth black hole solutions in shift symmetric kinetic gravity braiding. These conditions were used to obtain its most general stealth black hole solutions. 

It is not difficult to extend our strategy for building stealth solutions to the quartic sector of Horndeski theory (e.g., Gauss-Bonnet theories). For quintic Horndeski theories (e.g., theories with nonmininal derivative coupling with Einstein tensor), however, there are terms in the field equations which do not vanish for constant kinetic density and thus complicate a search for stealth solutions. This may be overcome by imposing additional restrictions on the potentials, but a detailed account of this is beyond the scope of the present work. 

Beyond these extensions, there are several other exciting avenues that may be pursued in future work. It goes without saying that any gravity theory must satisfy gravitational wave and cosmological constraints \cite{ligo_catalog, bh_physics_community, no_run_linder, ferreira2019cosmological}.
Going to something more specific, this paper only focused on vacuum solutions. In the presence of matter, scalar-tensor theories can possess screening mechanisms that effectively deactivate the scalar field at solar system scales \cite{screening_chakraborty, screening_brax, screening_schmidt, screening_davis, st_horndeski_vainshtein_dima, screening_mcmanus}. It would be interesting to isolate kinetic gravity braiding theories which accomodate \emph{both} stealth black hole solutions and screening mechanisms. The addition of an electric charge to a nonrotating black hole also appears straightforward, as is the introduction of an electromagnetic field for slowly rotating magnetized stealth black holes. Finally, an analysis of tensor perturbations of the stealth solutions is essential, as it would shed light on questions of stability and be of practical use for building GW models for space-based missions such as LISA. Clearly, a lot more effort is needed to completely understand gravitational perturbations in scalar-tensor theories \cite{st_bh_pt_de_rham, st_kerr_pt_charmousis, st_bh_pt_takahashi}.

\section*{Acknowledgements}
The authors are grateful to Hayato Motohashi and Masato Minamitsuji for helpful clarifications about their work. The authors are also grateful to Sean Fortuna for useful insights and feedback on the preliminary version of the manuscript. Throughout this work, the authors have fully exploited the `xAct' package \cite{xact} and its derivative `xCoba' \cite{xcoba}. This research is supported by the University of the Philippines OVPAA through Grant No.~OVPAA-BPhD-2016-13.

\bibliographystyle{apsrev4-1}

%

\end{document}